\documentstyle[12pt]{article}
\def\d{{\rm d}}
\def\l{\lambda }  \def\r{\varrho }
\def\o{\omega }    \def\t{\vartheta }

\newcommand{\beq}{\begin{equation}}
\newcommand{\beqar}{\begin{eqnarray}}
\newcommand{\eeq}{\end{equation}}
\newcommand{\eeqar}{\end{eqnarray}}
\begin{document}
\centerline{{\huge Shell Effects in Mesoscopic Systems}}
\medskip
\centerline{ R.G.~Nazmitdinov$^{\star}$ and W.D.~Heiss $^{\star \star}$}
\medskip
\centerline{{\sl
$^{\star}$ Bogoliubov Laboratory of Theoretical Physics}}
\centerline{{\sl
Joint Institute for Nuclear Research, 141980 Dubna, Russia}}
\centerline{{\sl
$^{\star \star}$ Centre for Nonlinear Studies and Department of Physics}}
\centerline{{\sl
University of the Witwatersrand, PO Wits 2050, Johannesburg, South Africa }}

\vspace{2cm}

A major property of confined Fermi system is the quantisation
of single particle motion. It leads to a bunching of levels
in the single particle spectrum, known as shells, and gives rise to magic
numbers in finite Fermi systems.
Consequently, a spherical symmetry leads to very strong shell effects
manifested in the stability of the noble gases, nuclei and metallic
clusters \cite{BM75,He}.
Recently the sequence of magic numbers of a two-dimensional
harmonic oscillator has been observed in
the addition energy for a vertical quantum dot,
i.e. in the energy needed to place the extra electron into
the dot \cite{Ta}. In contrast to nuclei and metallic clusters
the properties of quantum dots can be controlled by men.
The main aim of this talk to discuss the manifestation of shell
effects in the mean field approach in
different mesoscopic systems, i.e. in nuclei, metallic
clusters and quantum dots.

{\bf Periodic motion and higher multipoles in
nuclei and metallic clusters.}
When a spherical shell is
only partially filled, a breaking of spherical symmetry, resulting in an
energy gain, can give rise to a deformed equilibrium shape.
Super- and hyperdeformed nuclei are among the most fascinating examples
where deviations from the spherical shape are a
consequence of strong shell closures giving rise to largest level bunching
(largest degeneracy or lowest level density).
Nowadays it is recognized that the fine structure in the
mass spectra between magic numbers in metallic clusters
 could be explained via symmetry breaking
mechanisms similar to the situation in nuclear physics.
It is therefore accepted that cluster
deformations can exist, and it is actually confirmed at least for
clusters with $A\leq 40$ either by the Clemenger--Nilsson (CN) model
(introduced by Nilsson \cite{N} in nuclear physics and applied by Clemenger
\cite{Clem} for clusters) or by a self-consistent
Kohn--Sham density--functional method \cite{KS65} (KS) with deformed jellium
backgrounds \cite{EP}--\cite{Mon95}.
The need for multipole deformations higher than the quadrupole
in the mean field approach has
been recognized in nuclei and in metallic clusters in numerous calculations
to explain experimental data.  For instance, the octupole deformed shapes
still constitute an intriguing problem of
nuclear structure, experimental as well as theoretical
(see for review \cite{BN}).
The hexadecapole deformation is essential for the understanding
of equilibrium shapes and the
fission process of super- and hyperdeformed nuclei \cite{Dud}.
In the case of metallic clusters, the axial
hexadecapole deformation is important for the interpretation of
experimental data in simple metals \cite{HB}.
Inclusion of higher multipoles leads, however, to a nonitegrable
problem. In fact, the single particle motion turns out to be chaotic.
It is expected that an increasing strength of the corresponding multipole
deformation increases the amount of chaos in the classical potential
as quantified by an increasing Lyapunov exponent. Accordingly,
in the transition
>from ordered to chaotic motion the quantum numbers lose their significance,
and the system behaves like a viscous fluid \cite{Bl}. Therefore,
a disappearance of shell structure should be expected in the analogous quantum case.
However, recently the occurrence of shell structure has been reported for
many body systems like nuclei  and  metallic clusters \cite{Ar,HNR} at strong
octupole deformation. A major conclusion of \cite{HNR} is
that, albeit nonintegrable,
an octupole admixture to quadrupole oscillator potentials
leads, for some values of the octupole strength,
to a shell structure similar to a plain but more deformed quadrupole
potential. This result shows that there is a tendency of the system
to restore the original symmetry (degeneracy of levels)
 which is destroyed by the octupole term.

The equilibrium deformation is ultimately related to the behavior
of the single particle level density of the quantum Hamiltonian. According
to the semiclassical theory \cite{BB} the frequencies in the level
density oscillations of single particle spectra are determined
by the corresponding periods of classical closed orbits. The short periodic
orbits determine the gross shell structure, whereas contribution of longer
orbits give finer details. Analysis of shell structure phenomena in
nuclei in terms of classical orbits was started by the work of
Balian and Bloch
\cite{BB72} who studied the density of eigenmodes in a spherical cavity with
reflecting walls. The importance of deformed shapes for nuclei led to the
generalization of this analysis in considering a deformed ellipsoidal
well and a deformed
harmonic oscillator \cite{BM75,SM76}. Since these problems are
integrable, attention was given
to regular motion. One of the first attempts to analyze shell structure
phenomena and chaotic motion in a
quadrupole deformed diffuse cavity was given in \cite{Arv}.
Our approach is based on the connection between shell structure phenomena in
the quantum spectrum and ordered motion in the classical analogous case.
As it was mentioned above, shell structures in the quantum mechanical spectrum
are related to periodic orbits of the corresponding classical problem
\cite{BM75},\cite{BB72,SM76}. The periodic orbits are associated with
invariant tori of the Poincar\'e sections. If the classical
problem is chaotic, the invariant tori disintegrate or disappear \cite{Gu90},
and the shell structure of the quantum spectrum is affected by the degree
of chaos \cite{Arv}-\cite{fin}.
The classical treatment is based on the
secular perturbation theory \cite{lili}
and is particularly effective for a two
degrees of freedom system. The technique uses
action-angle variables of the unperturbed Hamiltonian and averages over
the faster phase. Usually, prior to such
an operation, a canonical transformation is necessary in order to remove the
initial resonance from the unperturbed Hamiltonian (in our case the
axial harmonic oscillator without higher multipoles).
In the new rotating frame, one
of the phases will only measure the slow variation of the variables about
the original resonance which now becomes a fixed elliptic point. The problem
is then treated by averaging over the remaining faster phase.
For super- or hyperdeformed system the frequency of the oscillation in the
$\r $-coordinate is twice as fast or faster than that in the $z$-coordinate.
Since the Hamilton function is periodic in both angles,
this approximation amounts to keeping the zero order term of its Fourier
expansion in the fast moving angle. The integration of the
Hamilton function over the angles can be done analytically, which makes
the approximation particularly attractive.
The full Hamilton function,
being a nonintegrable problem, is approximated by an effective
Hamilton function that is obtained by averaging over the fast oscillating
angle $\t _{\r }$ for the case considered.

An axial harmonic oscillator Hamiltonian (AHO)
which is deformed by arbitrary multipoles has the form
\begin{equation}
H={1\over 2m}(p_{\r }^2+p_z^2+{p_{\phi }^2\over \r ^2})+{\cal U}(r,\t )
\end{equation}
where
\begin{equation}
{\cal U}(r,\t ) =
\frac{m\omega^2}{2}\left( \r ^2 + \frac{z^2}{b^2} +
 r^2(\l  _3 P_3(\cos\t ) +
 \l  _4 P_4(\cos\t )+\ldots ) \right)
\label{eq:fact}
\end{equation}
with $r^2=\r ^2+z^2$, $\cos\t =z/r$. The $z$-component of the angular
momentum is denoted by $p_{\phi }$ and $P_k(\cos\t )$ is
the $k-$th order Legendre polynomial.
The parameter $b$
characterizes an oblate and a prolate shape for $0<b<1$ and $b>1$,
respectively. For a non-vanishing $\l _3,\l _4,\ldots $ the problem becomes a
nonintegrable two degrees of freedom system. Note that the only constant of
motion is the total energy.
Our interest is focused on the contribution of the octupole and
hexadecapole deformation in super- and hyperdeformed system.
Therefore, the general potential Eq.(\ref{eq:fact}) is reduced to the form
\beq
V(\r ,z)={m\over 2} \o ^2 (\r ^2+{z^2\over b^2}+\l _3 {2z^3-3z\r ^2 \over
\sqrt{\r ^2+z^2}}+\l _4 {8z^4 - 24 z^2\r ^2 + 3\r ^4\over z^2 + \r ^2}).
 \label{pot}  \eeq
The terms multiplied by
$\l _3$ and $\l _4$ give rise to octupole and hexadecapole deformations, the
respective terms are proportional to $r^2P_3(\cos \t )$ and
$r^2P_4(\cos \t )$.

The axial symmetry of the potential given in Eq.(\ref{pot})
guarantees conservation
of the $z$-component of the angular momentum denoted by $p_{\phi }$,
the discussion will be focussed on $p_{\phi }=0$ unless indicated otherwise.
After averaging and rewriting the action variables and the remaining
angle $\t _z$ in terms of the original momentum and coordinate values we
obtain \cite{fin}
\beq   H_{{\rm av}} = {p_{\r }^2+p_z^2\over 2m}+
{m\o ^2\over 2}\r ^2 + U_{{\rm eff}}(z)
\label{hav} \eeq
where
\beqar  U_{{\rm eff}}(z)=&&{m\o ^2\over 2}\bigl[{z^2\over b^2}+
\l _3\xi _{\r } ^2{ {\rm sign} (z)\over 2\pi }
\bigl(8{z^2\over \xi _{\r } ^2}K(-{\xi _{\r } ^2\over z^2})  -
3\pi {_2{\rm F}_1}({1\over 2},{3\over2},2;-{\xi _{\r } ^2\over z^2})\bigr)+
\nonumber \\
&&\l _4\bigl({3\over 2}\xi _{\r } ^2-27z^2+{35|z^3|\over
\sqrt{\xi _{\r } ^2+z^2}}\bigr)
\bigr].   \label{potef} \eeqar
and $\xi _{\r } ^2=2J_{\r }/(m\o )=2E_{\r }/(m\o ^2)$ which is
a constant within the approximation. Here $K$ is the first elliptic integral.
Note that the approximated Hamilton function is separable in the two
coordinates with the $\r $-motion being unperturbed. Consequently the
frequencies are $\o _{\r }=\o $ and $\o _z=2\pi / T_z $ with
\beq   T_z=\sqrt{2m}\int _{z_{{\rm min}}}^{z_{{\rm max}}}
   {\d z\over \sqrt{E-E_{\r }-U_{{\rm eff}}(z)}}  \label{peri} \eeq
The motion in the $z$-coordinate is different from the unperturbed motion
and its frequency depends on $\xi _{\r } $, i.e.~on the amount of the energy
residing in the $\r $-motion. We obtain for the winding number
\beq {\o _{\r }\over \o _z}={1\over 2}({1\over \sqrt{1/b^2-2\l _3+8\l _4}}
    +{1\over \sqrt{1/b^2+2\l _3+8\l _4}}).  \label{wind} \eeq
This result is exact for $H_{{\rm av}}$ when $\xi _{\r } = 0$
(all energy resides in
the $z$-motion), but by the previous argument it applies for large part
of phase space. Moreover, it serves as a useful and reliable guideline for
the full problem.

The oblate case ($b<1$) does not lend itself to the same approximation
procedure. As was reported in \cite{HNR} an octupole addition to an oblate
quadrupole potential gives no contribution to the zeroth order when
averaging over $\t _z$. If only a
hexadecapole term is added, the same procedure can be applied and we obtain
\cite{new}
\beq
 W_{{\rm eff}}(\r ) ={m\over 2}\o ^2 [\r ^2
 +\l _4 ({35|\r ^3|\over \sqrt{\r ^2+\xi _z^2}}-32\r ^2
    + 4\xi _z^2)]
     \label{veff}  \eeq
with $\xi _z^2= 2bE_z/(m\o ^2)$. In this case, we find for $\xi _z=0$,
for the winding number
\beq  {\o _{\r }\over \o _z}=b\sqrt{1+3\l _4}. \label{windob} \eeq
Note that the admixture of the hexadecapole term to the oblate potential
yields, quantum mechanically, an effective plain oblate case but with
less deformation. This is in contrast to adding an octupole to a prolate
potential where the effective deformation is increased.

The quantum-mechanical results are in line with the classical predictions.
We calculated the energy levels of the Hamiltonian
\beq
H = H_0 + \l _3 H_3 + \l _4 H_4
\label{hamqu}
\eeq
where the diagonal matrix $H_0$ comprises the axial harmonic oscillator,
and $H_3$ and $H_4$ are the octupole and
hexadecapole terms, respectively. The consistent calculation of the matrix
elements in a truncated basis is described in \cite{HNR}.

As a quantitative measure for shell structure we use the
Strutinsky type analysis (see for details \cite{HNR,fin}).
In accordance with the discussion above, the oblate superdeformed potential
produces chaos when the octupole term is switched on. The quantum spectrum
has the level statistics ascribed to quantum chaos \cite{US2}. The
hexadecapole term alone does gives rise to new shell structure;
for instance, for $b=2/5$ and
$\l _4\approx 0.1$ oblate superdeformation, i.e.~$\o _{\r }/\o _z=1/2$
was established. This is close to the value given by Eq.(\ref{windob}), the
difference is explained in \cite{new}.

There is remarkable agreement between the manifestation
of shell structure for values of the strength parameters $\l_3$, $\l_4$,
which coincide with the ones predicted by the classical perturbative
approach and give rise to stability islands on the Poincar\'e
surfaces of sections. According to the classical approach only even multipoles
can decouple the potential of Eq.(1) for an oblate deformation.
Therefore, shell structure can be supported for strongly oblate deformed
nuclei or clusters only by even multipoles.
A further interesting result is the mutual
cancellation of the octupole and hexadecapole contribution in the quadrupole
deformed system. The classical approach allows to determine the range of
parameters of $\l_3$ and $\l_4$ where the corresponding quantum mechanical
problem of the quadrupole + octupole + hexadecapole potential yields shell
structure resembling the plain quadrupole deformation \cite{fin}.
Along this curve the octupole deformation tends to increase
the effective prolate deformation whereas the hexadecapole term produces the
opposite effect. In fact, the hexadecapole term can stabilize the
octupole deformation in superdeformed systems, since the cancellation
curve attains values of $\l _3$ which are larger than its critical
value for $\l _4=0$. We may speculate that prolate
superdeformed nuclei with rather strong octupole deformation
could therefore exist. Finally we comment that even though the
quantum mechanical treatment shows a
certain degree of suppression of classical chaos, the occurrence of a new
shell structure which differs from the unperturbed case is clearly brought
about by the nonlinear character of the problem.

{\bf Filled shells in quantum dots.}
We choose the harmonic oscillator potential as the effective mean field for
the electrons in an isolated quantum dot. While the electron-electron
interaction is important for the explanation of certain
ground state properties like special values of angular momenta
of a quantum dot in a magnetic field \cite{MC},
a smooth and finite potential which admits bound states can for the
lowest few levels always be approximated by
the harmonic oscillator potential \cite{DE}. This has been confirmed
by direct determination of a new effective
{\it oscillator} frequency for two interacting electrons in an external
parabolic potential \cite{DN} and by calculations of the effective
single particle levels within the density-functional theory
for electron numbers $N\sim 100$ \cite{Sto}.
For small dot size and small number of electrons
the confinement energy becomes
prevalent over the Coulomb energy.
The experimental identification of the magic numbers of the two-dimensional
harmonic oscillator without a magnetic field \cite{Ta} is
the most convincing argument for the validity of this approximation.
The effect of an external homogeneous magnetic field can be calculated
exactly for a three dimensional (3D) harmonic oscillator potential
irrespective of the direction of the field \cite{HeNa97}.
Our discussion here is based upon the 2D version of the Hamiltonian
\cite{HeNa97} including spin degree of freedom. The magnetic field acts
perpendicular to the plane of motion, i.e.
 $H=\sum_{j=1}^A h_j$
with
\begin{equation}
h={1\over 2m^*}(\vec p-{e\over c}\vec A)^2+{m^*\over 2}(\omega _x^2 x^2
+\omega _y^2 y^2) + \mu^*\sigma _zB.           \label{ham}
\end{equation}
where $\vec A=[\vec r \times \vec B]/2, \, \vec B=(0,0,B)$ and
$\sigma _z$ is the Pauli matrix.
We do not take into account the effect of finite temperature; this is
appropriate for experiments which are performed at temperatures
$kT\ll \Delta$ with $\Delta $ being the mean level spacing. In the following
we use meV for the energy and Tesla for the magnetic field strength.
The effective mass
which determines the orbital magnetic moment for the electrons is chosen as
$m^*=0.067m_e$. It leads to $\mu _B^{{\rm eff}}=15\mu _B$ while the effective
spin magnetic moment is $\mu ^*=0.5\mu _B$.

Shell structure occurs whenever the ratio of the two eigenmodes
$\Omega _{\pm }$ of the Hamiltonian (\ref{ham}) (see Ref.\cite{HeNa97})
is a rational number with a small numerator and denominator.
If we start with a circular dot ($\omega_x=\omega_y$),
the shell structure is particularly pronounced if
the ratio is equal to one (for $B=0$) or two (for
$B\approx 1.23$ ) or three (for $B\approx 2.01$ ) and lesser
pronounced if the ratio is 3/2 (for $B=0.72$) or 5/2 (for $B=1.65$).
The values given here for $B$ depend on $m^*$ and $\omega _{x,y}$.
As a consequence, a material with an even smaller effective mass $m^*$ would
show these effects for a correspondingly smaller magnetic field.
The magic numbers (including spin) turn out, for $B=0$, to be the
usual sequence of the two dimensional isotropic oscillator, that is
$2,6,12,20,\ldots $ \cite{Ta}. For $B\approx 1.23$
we find a new shell structure {\em as if} the confining potential would be a
deformed harmonic oscillator without magnetic field. The magic numbers are
$2,4,8,12,18,24,\ldots $ which are just the numbers obtained from the two
dimensional oscillator with $\omega _>=2\omega _<$ ($\omega _>$ and
$\omega _<$ denote the larger and smaller value of the two frequencies).
Similarly, we get for $B\approx 2.01$ the magic numbers
$2,4,6,10,14,18,24,\ldots $ which corresponds to $\omega _>=3\omega _<$.
If we start from the outset with a deformed mean field, i.e.~if we choose,
say, $\omega _x=(1-\beta )\omega _y$ with $\beta >0$
two major effects are found:
(i) the degeneracies (shell structure) are lifted at $B=0$ depending on the
actual value of $\beta $, and (ii) the values for $B$ at which the new shell
structures occur are shifted to lower values.
The significance of this finding lies in the
restoration of shell structures by the magnetic field in an isolated
quantum dot that
does not give rise to magic numbers at zero field strength due to deformation.
We mention that the choice $\beta =0.5$ would shift the pattern found at
$B\approx 1.23$ to the value $B=0$.
It is the shell structure caused by the effective mean field
which produces the maxima that are observed experimentally in the addition
energy $\mu (A+1)- \mu (A) = E_{A+1}-E_{A}+ e^2/C$ \cite{Ta}.
Here $E_A$ is the single particle energy of the effective
mean field in quantum dots, $e^2/C$  is the electrostatic energy and
$\mu (A)$ is the
chemical potential. The electrostatic energy is much larger
than the difference $E_{A+1}-E_{A}$, however, it is the fluctuations (shell
effects) of the difference that matters, at least for small
quantum dots. The effect is similar to shell phenomena
in nuclear physics and for metallic clusters.
The analogy goes further in that, in an isolated small
quantum dot, the external magnetic
field acts like the rotation on a nucleus thus creating new shell structure;
in this way superdeformation (axis ratio 2:1) has been established
for rotating nuclei owing to the shell gaps in the single particle
spectrum.
We now focus on the special cases which give rise to pronounced shell
structure, that is when the ratio
$\Omega_+/\Omega_-=k=1,2,3,\ldots $,
and analyse in detail the
circular shape ($\omega_x=\omega_y=\omega_0$).
We find for the magnetization
\begin{equation}
\label{mag1}
M= \mu_B^{{\rm eff}}(1-{\omega_L \over \Omega})
({\sum}_{-}-k {\sum}_{+})- \mu^*<S_z>
\end{equation}
with  $\sum_{\pm } = \sum _j^A(n_{\pm }+1/2)_j$ \cite{HeNa97},
$\omega_L= \alpha B = \frac{|e|}{2mc} B$ and
$\mu_B^{{\rm eff}}=\hbar \alpha$ being the effective Bohr magneton using
the effective mass $m^*$.
The eigenmodes are $\Omega_{\pm} = (\Omega \pm \omega_L )$ with
$\Omega = \sqrt{\omega_0^2+\omega_L^2}$ \cite{Fock}.

For completely filled shell $<S_z>=0$, since, for the magnetic field
strenghts considered here, the spin orientations cancel each other.
>From the orbital motion we obtain for the susceptibility
\begin{equation}
\label{suc}
\chi=dM/dB=-\frac{{\mu_B^{{\rm eff}}}^2}{\hbar \Omega} (\frac{\omega_0}{\Omega})^2
({\sum}_{+}+{\sum}_{-})
\end{equation}
It follows from Eq.(\ref{suc}) that, for a completely filled shell,
the magnetization owing to the orbital motion leads to
diamagnetic behaviour. For zero magnetic field ($k=1$) the
 system is paramagnetic and the magnetization vanishes
($\sum_- = \sum_+$). The value $k=2$ is attained at $B\approx 1.23$.
When calculating $\sum_-$ and $\sum_+$  we now have to
distinguish between the cases, where the shell number $N$ of the last filled
shell is even or odd. For the last filled shell number even one finds
\begin{eqnarray}
{\sum}_+ = {1\over 12}(N+2)[(N+2)^2+2]\\
{\sum}_- = {1\over 6}(N+1)(N+2)(N+3)
\end{eqnarray}
and $M=-\mu_B^{{\rm eff}}(1-{\omega_L / \Omega})(N+2)/2.$

For the last filled shell number odd one finds
\begin{equation}
{\sum}_+ = {1\over 2}{\sum}_- = {1\over 12}(N+1)(N+2)(N+3)\\
\end{equation}
and $M=0.$
Therefore, if $\Omega_+/\Omega_-=2$, the orbital magnetization vanishes
 for the magic numbers
$4,12,24,\ldots $ while it leads to diamagnetism for the magic numbers
$2,8,18,\ldots $. A similar picture is obtained for ${\Omega_+}/\Omega{-}=3$
which happens at $B\approx 2.01$: for each third filled shell number
(magic numbers $6,18,\ldots $) the magnetization is zero.
Since the results presented are due to shell effects, they do not depend
on the assumption $\omega _x/\omega _y=1$ which was made to facilitate the
discussion. The crucial point is the relation
$\Omega_+/\Omega_-=k=1,2,3,\ldots $ which can be obtained for a variety
of combinations of the magnetic field strength and the ratio
$\omega _x/\omega _y$.
Whenever the appropriate combination of field strength and deformation is
chosen to yield, say, $k=2$, our findings apply.

In conclusion: our analysis for nuclei, metallic clusters and quantum dots
 is based on
a rather simple model, and it is true that
a realistic single particle spectrum is poorly reproduced by the
the harmonic oscillator. However,
the lucidity and transparence of the model in describing
the phenomena of extreme deformations for nuclei and metallic clusters
is superior to any realistic model.
The shell structure effects observed for the addition energy of a small
isolated quantum dot provide a reliable basis for use the 2D version of
the harmonic oscillator Hamiltonian. At certain values of the magnetic
field strength shell structures appear in a quantum dot, also in cases where
deformation does not give rise to magic numbers at zero field strength.
Measurements of the magnetic susceptibility are expected to reflect the
properties of the single particle spectrum and should display characteristic
patterns depending on the particle number.
At certain values of the magnetic field and electron numbers
the orbital magnetization vanishes due to shell closure
in the quantum dot.
This property could be of interest in applications
because it enables control of the electron number in small isolated
quantum dots.

\end{document}